\documentclass[12pt]{article}
\usepackage{amssymb} 
\usepackage{epsfig}
\usepackage{float}
\newcommand{\be}{\begin{equation}}
\newcommand{\ee}{\end{equation}}
\newcommand{\bea}{\begin{eqnarray}}
\newcommand{\eea}{\end{eqnarray}}

\begin{document} 

\begin{center}
\bf{A REMARK ON NEUTRINO OSCILLATIONS AND TIME-ENERGY UNCERTAINTY RELATION}

\end{center}

\begin{center}
S. M. Bilenky 
\end{center}

\begin{center}
{\em  Joint Institute
for Nuclear Research, Dubna, R-141980, Russia, and\\
Scuola Internazionale Superiore di Studi Avanzati, 
I-34014 Trieste, Italy.}
\end{center}

\begin{abstract}
Neutrino oscillations are discussed from the point of view of the time-energy uncertainty relation.

\end{abstract} 

\section{Introduction}

Strong model independent evidence in favor of  neutrino oscillations was obtained  recently in the 
atmospheric Super-Kamiokande \cite{SK}, solar SNO \cite{SNO}, reactor KamLAND 
\cite{kamland} and accelerator 
K2K \cite{K2K} neutrino experiments.  

All existing neutrino oscillation data, with the exception 
of the data of the LSND experiment \cite{LSND},  which presumably will be checked in the near future by 
the running MiniBooNE experiment \cite{ MiniB}, can be
described if we assume that three massive and mixed neutrinos exist in nature.

The data of the atmospheric SK and K2K experiments are perfectly described 
if we assume that $\nu_{\mu}$ ($\bar\nu_{\mu}$) survival probability has the standard two-neutrino form
\be
P(\nu_{\mu} \to\nu_{\mu}) =P(\bar\nu_{\mu} \to\bar\nu_{\mu})=1
- \frac{1}{2}\,\sin^{2}2\,\theta_{23} \,(1 - \cos\frac{\Delta m_{32}^2\,L }{ 2 E }),
\label{1}
\ee
where $E$ is the neutrino energy, $L$ is the distance between neutrino source and neutrino detector and
$\Delta m^2_{ ik }=m^2_{ i }-m^2_{ k }$ ($m_{i}, m_{k}$ are neutrino masses, $m_{1}< m_{2} <m_{3}$)

From analysis of the data of the SK experiment the following best-fit values of the oscillation parameters were found \cite{SK}:
\be
\sin^{2}2\,\theta_{23}=1;\,~~\Delta m_{32}^2= 2\cdot 10^{-3}\, \rm{eV}^{2}\,~~(\chi^2/\rm{dof}=170.8/172).
\label{2}
\ee
The data of the reactor KamLAND experiment are well described if we assume that 
oscillations of the reactor $\bar\nu_{e}$'s are driven by $\Delta m_{21}^2$ and
$\bar\nu_{e}$ survival probability has the two-neutrino form
\be
P(\bar\nu_{e} \to\bar\nu_{e}) =1
- \frac{1}{2}\,\sin^{2}2\,\theta_{12} \,(1 - \cos\frac{\Delta m_{21}^2\,L }{ 2 E }).
\label{3}
\ee
From the analysis of the  data of the KamLAND and solar neutrino experiments 
 (assuming CPT invariance) for neutrino oscillation 
parameters the following values were found \cite{kamland}
\be
\tan^{2}\theta_{12}=0.40^{+0.10}_{-0.07};\,~~\Delta m_{21}^2 =7.9^{+0.6}_{-0.5}\cdot 10^{-3}\, \rm{eV}^{2}.
\label{4}
\ee
Let us notice that there are the following two reasons, why existing neutrino oscillations data 
are described by the two-neutrino expressions (\ref{1}) and (\ref{3}) 
(see, for example, \cite{BGG}) :
\begin{enumerate}
\item
\be
\Delta m^{2}_{21}\ll \Delta m^{2}_{32}.
\label{5}
\ee
\item
\be
|U_{e3}| \ll 1
\label{6}
\ee
\end{enumerate}
This last inequality follows from the negative result of the reactor CHOOZ experiment \cite{CHOOZ}.

In spite of the strong evidence in favor of  neutrino oscillations, obtained in many neutrino experiments, 
the basics of this phenomenon is 
still a subject of intensive discussions and debates (see recent review \cite{Giunti} and 86 references therein). 
We  will add to 
these discussions some points which to our knowledge were not considered.

\section{Time-energy uncertainty relation is a condition to observe neutrino oscillations}

Our discussion will be based on the following ascertains:
\begin{enumerate}
\item
Quantum field theory is a natural framework for the consideration of 
the transitions between different flavor neutrinos.
\item
The evolution equation in the quantum field theory is the Schrodinger equation 
\be
i\,\frac{\partial |\Psi(t)\rangle}{\partial t} = H\, |\Psi(t)\rangle.
\label{7}
\ee
\item
Energies of neutrinos in neutrino experiments ($\gtrsim$ MeV) are much larger than neutrino masses
($\lesssim$ (1-2) eV).
At such energies neutrino masses can be neglected in matrix elements of neutrino production and 
detection processes and
the states of flavor neutrinos $\nu_{e}, \nu_{\mu}, \nu_{\tau}$   
(particles which take part in standard CC and NC weak processes) are {\em mixed states} which are described by the vectors (see, for example, \cite{BilG}):
\be
|\nu_{l}\rangle
= \sum_{i=1}^{3}|\nu_i\rangle\,~  U^{*}_{l i};\,~~~(l=e, \mu, \tau).
\label{8}
\ee
Here $U$ is an unitary 
3$\times$3 PMNS \cite{BP,MNS} mixing matrix,  $|\nu_i\rangle$  
is the states of left-handed neutrino 
with momentum
$\vec{p}$ and mass $m_{i}$.\footnote{The state of a particle in the quantum field theory is characterized 
by momentum, helicity, mass and internal quantum numbers. The state of mixed flavor neutrino 
is characterized by momentum, helicity, masses $m_{i}$ and elements of the mixing matrix. In order to 
calculate observable quantities it is necessary to average over neutrino spectrum, resolution of detector 
etc.}

\end{enumerate}

It follows from 2. and 3. that  
if flavor neutrino $\nu_{l}$ with momentum $\vec{p} $ was produced 
 at the initial time $t_{0}$ the state of neutrino at the time  $t$ 
is given by
\be
|\nu_{l}\rangle_{t}
= \sum_{i=1}^{3}|\nu_i\rangle\, e^{-iE_{i}\,(t-t_{0})}\,U_{l i}^* ,
\label{9}
\ee
where
\be
E_{i}=\sqrt{p^{2}+m^{2}_{i}} \simeq p +\frac{m^{2}_{i} }{ 2 p }.
\label{10}
\ee
Thus, the state of neutrino at the time $t>t_{0}$ is described 
by {\em 
a superposition of the stationary states}. 

It is a general feature of the quantum theory that for such states the time-energy 
uncertainty relation
\footnote{As it is well known,  the time-energy uncertainty relation
and Heisenberg uncertainty relations 
$\Delta p\, \Delta x \gtrsim 1$ etc have completely different origin. 
In fact, time 
in the quantum theory is a parameter, there is no operator which corresponds to time. In 
the Heisenberg uncertainty relations enter uncertainties of
two observables
operators of which satisfy canonical commutation relations.} 
\be
\Delta E\, \Delta t \gtrsim 1.
\label{11}
\ee
takes place (see, for example \cite{Sakurai}). Here $\Delta t $ is a characteristic time interval 
during which significant changes in the system happen.

In the case of the mixing of two massive neutrinos
we have 
\be
\Delta E =E_{2}- E_{1}\simeq \frac{\Delta m^{2}_{21}}{2\,p};\,~~\Delta t =t-t_{0}.
\label{12}
\ee

Neutrinos are detected via the observation of CC and NC weak processes.
For the state $|\nu_{l}\rangle_{t}$ we obtain  
\be
|\nu_{l}\rangle_{t}
= \sum_{l'=e,\mu,\tau}|\nu_{l'}\rangle\, \mathcal{A}_{\nu_{l'};\nu_{l}}(t-t_{0}).
\label{13}
\ee
Here
\be
\mathcal{A}_{\nu_{l'};\nu_{l}}(t-t_{0})=\sum_{i=1}^{3} U_{l' i} \,e^{i\,E_{i}\, (t -t_{0})}\,U^{*}_{l  i} 
=e^{i\,E_{1}\, (t -t_{0})}\,~
\sum_{i=1}^{3} U_{l' i} \,e^{-i\,\frac{ \Delta m^2_{ i1 }}{ 2 p}\, (t -t_{0}) }\,U^{*}_{l  i}.
\label{14}
\ee
is the amplitude of $\nu_{l}\to \nu_{l'}$ transition in vacuum during the time interval 
$(t -t_{0})$. 

Taking into account the unitarity of the mixing matrix $U$, for the $\nu_{l}\to \nu_{l'}$ transition 
probability we obtain the following 
standard expression 
\be
P(\nu_{l} \to\nu_{l'}) 
=|\delta_{l' l}+ 
\sum_{i=2,3} U_{l' i} \,(e^{\frac{ \Delta m^2_{ i1 }}{ 2 p }\, L }-1)
U^{*}_{l i}|^{2}, 
\label{15}
\ee
where the distance between neutrino source 
and neutrino detector $L$ is given by
\footnote{This relation was used (and checked) in the K2K 
experiment \cite{K2K}. 
In order to produce neutrino beam protons from KEK accelerator are extracted 
in  1.1 $\mu sec$ spills 
every 2.2 $sec$.
Events which satisfy the criteria $-0.2\leq\Delta t \leq 1.3 \, \mu sec$  are selected in the 
experiment. Here
$\Delta t = t_{SK}-t_{KEK}-t_{TOF}$, $t_{KEK}$ is the measured time of the production of neutrinos at KEK, 
$t_{SK}$ is the  measured time of  the detection of neutrinos in the Super-Kamiokande detector,
$t_{TOF}=L/c\simeq 0.83\cdot 10^{3}\,\mu sec $}
\be
L\simeq (t-t_{0}).
\label{16}
\ee
In the  case of the flavor neutrino transitions, driven by one neutrino mass-squared difference, neutrino oscillations can be observed at such distances L 
at which 
the following inequality 
\be
\frac{\Delta m^2}{2\,p} \,L \gtrsim 1.
\label{17}
\ee
 is satisfied \cite{BilPont}.

It is obvious from (\ref{12}), (\ref{16}) and (\ref{17}) that the condition of the observation of neutrino oscillations is  time-energy uncertainty relation. 
The characteristic time of transitions between different flavor neutrinos is determined by the oscillation 
time (length) given by the equation
\be
T_{0}\simeq L_{0}=4\,\pi\, \frac{ p }{\Delta m^2}
\label{18}
\ee
\section{Flavor neutrino states and translations}
Let us consider 
translations
\be
x'=x+a,
\label{19}
\ee
where $x$ and $x'$ are coordinates of the same point in two different systems  and $a$ is 
a constant arbitrary vector. In the case of 
the invariance under translations the states of  
{\em the same physical system} in these two reference systems are connected by the relation
\be
|\Psi \rangle' = e^{i\,P\, a }\, |\Psi \rangle,
\label{20}
\ee
and an operator 
$O(x)$ satisfies the relation
\be
O(x+a)=  e^{i\,P\, a }\,O(x)\, e^{-i\,P\, a }
\label{21}
\ee
Here $P$ is the operator of the total momentum. 

If the system has definite total momentum $p$,
the states $|\Psi \rangle'$ and  $|\Psi \rangle$ differ by the phase factor:
\be
|\Psi \rangle' = e^{i\,p\, a }\, |\Psi \rangle .
\label{22}
\ee
This relation provides conservation of the total momentum.

Let us apply now the operator of the translations $e^{i\,P\,a}$ to the flavor state 
$|\nu_{l}\rangle$. 
We have
\be
|\nu_{l}\rangle'=    e^{i\,P\, a }\, |\nu_{l}\rangle= e^{-i\,\vec{p}\,\vec{a}}\,\sum_{i}
 |\nu_i\rangle\,e^{i\,E_{i}\,a}\, U_{l i}^*=  e^{-i\,\vec{p}\,\vec{a}}\,\sum_{l'} 
|\nu_{l'}\rangle \,\sum_{i}U_{l' i}e^{i\,E_{i}\,a}\,U_{l i}^*
\label{23}
\ee
Thus, the vectors $|\nu_{l}\rangle'$ and $|\nu_{l}\rangle$ describe {\em different
states}. We come to the conclusion that in the case of the mixed flavor states
there is no invariance under translations. This 
means that in   transitions between different flavor neutrinos energy in principle is not conserved.
Non conservation of energy in neutrino oscillations is obviously connected with finite time between
neutrino production and neutrino detection and with the time-energy 
uncertainty relation. It has no any practical manifestations except neutrino oscillations.
\section{Conclusions}
From our point of view neutrino oscillations, observed in many neutrino experiments, is due to the fact that {\em coherent} flavor neutrino states 
are produced and detected in weak interaction processes. 
The state of neutrino at the time $t$ is {\em a superposition of the stationary 
states}. For such states exist characteristic time during which the flavor content of the state is significantly 
changed. This characteristic oscillation time satisfy the classical time-energy uncertainty relation.

There exist a claim (see, for example, \cite{Lipkin,Field}) that the arguments of cosines in  
the transition probabilities
 are two times larger than in (\ref{1}), (\ref{3}) and other expressions.
There is no way to obtain such (from our point of view erroneous) result in the framework of the field-theoretical 
approach presented here.

It is my pleasure to acknowledge 
the Italian program
``Rientro dei cervelli'' for the support.

\end{document}